\documentclass[12pt,a4paper]{article}

\usepackage[utf8]{inputenc}
\usepackage{amsmath,amssymb,amsfonts}
\usepackage{bm}
\usepackage{geometry}
\usepackage{cite}
\usepackage{hyperref}

\numberwithin{equation}{section}

\providecommand{\keywords}[1]{
\small
\noindent\textbf{Keywords: } #1
}

\title{
Hawking Temperatures of Dynamical Black Holes from the RVB--Residue Method:
Vaidya and Kinnersley Geometries
}

\author{
    Wen-Xiang Chen\\
    School of Electronic Information\\
    Guangzhou City University of Technology\\
    \texttt{wxchen4277@qq.com}
}

\date{}

\begin{document}

\maketitle

\begin{abstract}
We formulate a local horizon-residue prescription, motivated by the
Robson--Villari--Biancalana (RVB) construction, for non-stationary black holes.
The claim is deliberately local: it is not a construction of a global Euclidean
equilibrium state for a dynamical spacetime.  For a spherically symmetric metric
$ds^{2}=-C(v,r)dv^{2}+2dv\,dr+r^{2}d\Omega^{2}$ possessing a non-degenerate
future outer trapping horizon, we show step by step that the outgoing
Hamilton--Jacobi action contains the same simple pole as $1/C$.  Its residue is
$1/(2\kappa_{\rm H})$, where $\kappa_{\rm H}$ is the Kodama--Hayward surface
gravity, and the tunnelling rate gives
$T_{\rm H}=\kappa_{\rm H}/(2\pi)$.  This establishes the physical origin and the
precise domain of the residue prescription rather than inferring it only from
examples.  The Vaidya calculation then provides an exact check at the trapping
horizon; a separate event-horizon calculation is retained only as a clearly
labelled quasi-stationary comparison.  For the non-spherical Kinnersley
geometry, we derive the inverse metric, null-horizon equation, generalized
tortoise normalization, and the local radial function explicitly.  The latter
has slope $2\kappa_{\rm K}$ at the horizon and hence reproduces the known
point-dependent temperature.  The prescription requires a holomorphic simple
zero and a non-vanishing normalization denominator; degenerate or non-analytic
horizons lie outside the result proved here.
\end{abstract}


\section{Introduction}

The Hawking temperature of a stationary black hole is a central result of
semiclassical gravity. It originates from Hawking's discovery of particle
creation by black holes and is closely related to Bekenstein's entropy-area
relation and the four laws of black-hole mechanics
\cite{Hawking1974,Hawking1975,Bekenstein1973,Bardeen1973,GibbonsHawking1977}.
The standard expression is
\begin{equation}
T_{\mathrm{H}}=\frac{\kappa}{2\pi},
\end{equation}
where $\kappa$ is the surface gravity.

For a static and spherically symmetric metric of the form
\begin{equation}
ds^{2}=-F(r)dt^{2}+\frac{dr^{2}}{F(r)}+r^{2}d\Omega^{2},
\end{equation}
the Killing horizon is determined by
\begin{equation}
F(r_{\mathrm{H}})=0.
\end{equation}
The corresponding surface gravity is
\begin{equation}
\kappa=\frac{1}{2}F'(r_{\mathrm{H}}).
\end{equation}
Therefore, the Hawking temperature becomes
\begin{equation}
T_{\mathrm{H}}=\frac{F'(r_{\mathrm{H}})}{4\pi}.
\end{equation}

The Robson--Villari--Biancalana method provides a topological interpretation
of the Hawking temperature by relating the Euclideanized two-dimensional
black-hole geometry to the Euler characteristic of the corresponding manifold
\cite{Robson2019,RobsonAdS2019,ZhangWeiLiu2020}. In its residue-based extension,
the temperature is extracted from the complex near-horizon structure of the
radial horizon function. This idea has been applied by Chen to black holes in
modified gravity, especially in $f(Q)$ gravity, where extra correction terms may
be interpreted as residue contributions from complex contour integrals
\cite{Chen2025CJP,Chen2026Entropy,Chen2026Dirac}.

The original RVB formula may be formally written as
\begin{equation}
T_{\mathrm{RVB}}
=
\frac{1}{4\pi\chi}
\sum_{j}
\int_{r_{\mathrm{H},j}}
\sqrt{g}\,R\,dr,
\end{equation}
where $\chi$ is the Euler characteristic, $R$ is the Ricci scalar of the
two-dimensional Euclidean sector, and the summation is taken over the relevant
horizon points \cite{Robson2019}. For a single simple horizon, this expression
reduces to the ordinary surface-gravity result.

In the RVB--residue interpretation, the radial horizon function is analytically
continued into the complex plane. Let
\begin{equation}
z=r-r_{\mathrm{H}}\in\mathbb{C}.
\end{equation}
For a simple horizon, the local radial function admits the expansion
\begin{equation}
\mathcal{P}(z)
=
p_{1}z+p_{2}z^{2}+p_{3}z^{3}+\mathcal{O}(z^{4}),
\end{equation}
where
\begin{equation}
p_{1}
=
\left.
\frac{\partial \mathcal{P}}{\partial r}
\right|_{r=r_{\mathrm{H}}}.
\end{equation}
The inverse horizon function then has a simple pole:
\begin{equation}
\frac{1}{\mathcal{P}(z)}
=
\frac{1}{p_{1}}\frac{1}{z}
-\frac{p_{2}}{p_{1}^{2}}
+\mathcal{O}(z).
\end{equation}
Therefore,
\begin{equation}
\operatorname{Res}_{z=0}
\left[
\frac{1}{\mathcal{P}(z)}
\right]
=
\frac{1}{p_{1}}.
\end{equation}
The RVB--residue temperature is then defined by
\begin{equation}
T_{\mathrm{RVB-res}}
=
\frac{1}{4\pi}
\left(
\operatorname{Res}_{z=0}
\left[
\frac{1}{\mathcal{P}(z)}
\right]
\right)^{-1}.
\end{equation}
Thus,
\begin{equation}
T_{\mathrm{RVB-res}}
=
\frac{p_{1}}{4\pi}.
\end{equation}
If
\begin{equation}
p_{1}=2\kappa,
\end{equation}
then
\begin{equation}
T_{\mathrm{RVB-res}}
=
\frac{\kappa}{2\pi}.
\end{equation}

For dynamical black holes, the above construction must be interpreted locally.
A non-stationary black hole usually has no global timelike Killing vector, so
one must use local notions such as trapping horizons, Kodama--Hayward surface
gravity, Hamilton--Jacobi tunnelling, or generalized tortoise-coordinate
reductions \cite{Kodama1980,Hayward1994,Hayward1998,AshtekarKrishnan2004,
NielsenVisser2006,DiCriscienzo2010,Vanzo2011,Faraoni2015}.

The precise question addressed here is narrower than a global topological
extension: under what local and verifiable conditions does the reciprocal of a
dynamical horizon function have a residue that reproduces the semiclassical
temperature?  We prove the following restricted statement.  If (i) the
horizon is a simple zero of a locally holomorphic radial function,
(ii) the function is normalized by the regular near-horizon wave equation, and
(iii) an approximately thermal interpretation is meaningful over the emission
time scale, then the inverse function has residue $1/(2\kappa)$ and yields
$T=\kappa/(2\pi)$.  The topological RVB formula motivates the horizon-local
boundary contribution, while the Hamilton--Jacobi pole supplies its local
physical interpretation.  We do not claim that a generic non-stationary
spacetime possesses a global Euclidean section or a global equilibrium
temperature.

The specific contribution of the paper is therefore threefold: a general
derivation for the Eddington--Finkelstein class, an exact Vaidya check with an
explicit distinction between trapping and event horizons, and a derivation of
the normalized Kinnersley radial function rather than an assumption of its
linear coefficient.

\section{Local horizon-residue proposition and its physical origin}

We use units $G=c=\hbar=k_{\rm B}=1$ and signature $(-,+,+,+)$ in the
spherically symmetric part of the discussion.  A convenient regular chart is
the advanced Eddington--Finkelstein form
\begin{equation}
ds^{2}=-C(v,r)dv^{2}+2dv\,dr+r^{2}d\Omega^{2}.
\label{eq:generalEF}
\end{equation}
The two-dimensional normal metric and its inverse are, step by step,
\begin{equation}
(h_{ab})=
\begin{pmatrix}
-C&1\\
1&0
\end{pmatrix},
\qquad
\det h=-1,
\end{equation}
and
\begin{equation}
(h^{ab})=
\begin{pmatrix}
0&1\\
1&C
\end{pmatrix},
\qquad
h^{ac}h_{cb}=\delta^{a}{}_{b}.
\end{equation}
Because $r$ is the areal radius,
\begin{equation}
h^{ab}\nabla_{a}r\nabla_{b}r=h^{rr}=C(v,r).
\end{equation}
The relevant surface for a dynamical Hawking process is not chosen
arbitrarily. In the absence of a Killing horizon, the invariant replacement is
the future outer trapping horizon, because it is the zero of the Kodama energy
denominator in the Hamilton--Jacobi equation. We therefore select the local
emission surface to be a non-degenerate future outer trapping horizon,
\begin{equation}
C(v,r_{\rm H}(v))=0,
\qquad
C_{,r}(v,r_{\rm H})>0.
\label{eq:selectioncriterion}
\end{equation}
The second inequality fixes the outer branch and ensures a positive local
surface gravity.  It also states that the zero is simple.

The Kodama--Hayward surface gravity is
\begin{equation}
\kappa_{\rm H}=\frac12\Box_{h}r\big|_{\rm H}.
\end{equation}
No intermediate step is needed implicitly here.  Since
$\sqrt{-h}=1$ and $\partial_{b}r=\delta^{r}_{b}$,
\begin{align}
\Box_{h}r
&=\frac{1}{\sqrt{-h}}
\partial_{a}\!\left(\sqrt{-h}\,h^{ab}\partial_{b}r\right)\\
&=\partial_{a}h^{ar}\\
&=\partial_{v}h^{vr}+\partial_{r}h^{rr}\\
&=\partial_{v}(1)+\partial_{r}C(v,r)\\
&=C_{,r}(v,r).
\end{align}
Consequently,
\begin{equation}
\kappa_{\rm H}=\frac12 C_{,r}(v,r_{\rm H}),
\qquad
T_{\rm H}=\frac{C_{,r}(v,r_{\rm H})}{4\pi}.
\label{eq:haywardtemperature}
\end{equation}

\subsection{Existence and evaluation of the residue}

Fix a horizon event $(v_{0},r_{0})$, with
$r_{0}=r_{\rm H}(v_{0})$, and complexify only the transverse radial
coordinate,
\begin{equation}
z=r-r_{0}\in\mathbb{C}.
\end{equation}
The residue statement requires more than a real Taylor polynomial: we assume
that $C(v_{0},r_{0}+z)$ has a holomorphic continuation in a disk
$|z|<\delta_{0}$ and that no other zero lies in the contour used below.  With
\begin{equation}
C_{n}=\frac{1}{n!}
\left.\frac{\partial^{n}C(v_{0},r)}{\partial r^{n}}\right|_{r=r_{0}},
\end{equation}
conditions \eqref{eq:selectioncriterion} give $C_{0}=0$ and $C_{1}\ne0$.
Thus
\begin{equation}
C(v_{0},r_{0}+z)=C_{1}z+C_{2}z^{2}+C_{3}z^{3}+\mathcal{O}(z^{4}).
\label{eq:localTaylor}
\end{equation}
To invert the series without suppressing a step, write
\begin{equation}
\frac{1}{C(v_{0},r_{0}+z)}
=\frac{a_{-1}}{z}+a_{0}+a_{1}z+\mathcal{O}(z^{2}).
\end{equation}
Multiplication by \eqref{eq:localTaylor} gives
\begin{align}
1
&=(C_{1}z+C_{2}z^{2}+\cdots)
\left(\frac{a_{-1}}{z}+a_{0}+a_{1}z+\cdots\right)\\
&=C_{1}a_{-1}
+(C_{1}a_{0}+C_{2}a_{-1})z+\mathcal{O}(z^{2}).
\end{align}
Equating equal powers of $z$ yields
\begin{equation}
C_{1}a_{-1}=1,
\qquad
C_{1}a_{0}+C_{2}a_{-1}=0,
\end{equation}
and hence
\begin{equation}
a_{-1}=\frac{1}{C_{1}},
\qquad
a_{0}=-\frac{C_{2}}{C_{1}^{2}}.
\end{equation}
Therefore
\begin{equation}
\operatorname*{Res}_{z=0}\frac{1}{C(v_{0},r_{0}+z)}
=\frac{1}{C_{1}}
=\frac{1}{2\kappa_{\rm H}},
\label{eq:residuekappa}
\end{equation}
where the final equality follows from \eqref{eq:haywardtemperature}.  The
contour form is
\begin{equation}
\frac{1}{2\pi i}\oint_{|z|=\epsilon}
\frac{dz}{C(v_{0},r_{0}+z)}=\frac{1}{2\kappa_{\rm H}},
\qquad 0<\epsilon<\delta_{0}.
\end{equation}
It follows directly that
\begin{equation}
T_{\rm res}
=\frac{1}{4\pi}
\left(\operatorname*{Res}_{z=0}\frac{1}{C}\right)^{-1}
=\frac{\kappa_{\rm H}}{2\pi}.
\label{eq:residuetemperature}
\end{equation}

\subsection{Why the pole represents a Hawking temperature}

The physical content of \eqref{eq:residuetemperature} is exposed by an
independent Hamilton--Jacobi calculation
\cite{DiCriscienzo2010,Vanzo2011}.  At leading WKB order, a massless radial
mode with action $I$ obeys
\begin{equation}
h^{ab}\partial_{a}I\,\partial_{b}I=0.
\end{equation}
Using the inverse metric above gives
\begin{equation}
2(\partial_{v}I)(\partial_{r}I)
+C(v,r)(\partial_{r}I)^{2}=0.
\end{equation}
The Kodama vector in the present gauge is $K^{a}=(1,0)$, so the invariant
Kodama energy is
\begin{equation}
\omega=-K^{a}\partial_{a}I=-\partial_{v}I.
\end{equation}
Substitution produces
\begin{equation}
-2\omega\,\partial_{r}I+C(\partial_{r}I)^{2}=0,
\end{equation}
with the two roots
\begin{equation}
\partial_{r}I_{\rm in}=0,
\qquad
\partial_{r}I_{\rm out}=\frac{2\omega}{C(v,r)}.
\end{equation}
Only the outgoing root contains the horizon pole.  Equations
\eqref{eq:localTaylor} and \eqref{eq:residuekappa} give
\begin{equation}
\partial_{r}I_{\rm out}
=\frac{2\omega}{C_{1}z}+\mathcal{O}(1)
=\frac{\omega}{\kappa_{\rm H}}\frac{1}{z}+\mathcal{O}(1).
\end{equation}
With the causal prescription $z\rightarrow z-i0$, crossing the pole contributes
\begin{equation}
\operatorname{Im}I_{\rm out}
=\frac{2\pi\omega}{C_{1}}
=\frac{\pi\omega}{\kappa_{\rm H}}.
\end{equation}
The leading tunnelling probability is consequently
\begin{equation}
\Gamma\propto
\exp[-2\operatorname{Im}I_{\rm out}]
=\exp\!\left(-\frac{2\pi\omega}{\kappa_{\rm H}}\right)
=\exp\!\left(-\frac{\omega}{T_{\rm H}}\right),
\end{equation}
which independently fixes
\begin{equation}
T_{\rm H}=\frac{\kappa_{\rm H}}{2\pi}.
\end{equation}
The residue is therefore not an arbitrary algebraic assignment: it is the
coefficient of the same simple pole that controls the imaginary part of the
outgoing semiclassical action.

The horizon-selection criterion is now explicit.  In the local spherical
problem, the physically relevant surface is the zero of the outgoing-mode
denominator $C$, equivalently the future outer trapping horizon
\eqref{eq:selectioncriterion}.  A global event horizon can be considered only
after it has been determined from global boundary data, and its normalization
must be stated separately.  Finally, an approximately Planckian instantaneous
spectrum requires an adiabaticity condition such as
\begin{equation}
\epsilon_{\rm ad}
=\frac{|K^{a}\nabla_{a}\kappa_{\rm H}|}{\kappa_{\rm H}^{2}}\ll1.
\label{eq:adiabaticity}
\end{equation}
The local residue and surface gravity remain defined when this inequality
fails, but interpreting them as an equilibrium temperature would then be too
strong.

\section{Vaidya Black Hole}

This section is an exact check of the general proposition, not the basis from
which it is inferred.  The local surface selected by
\eqref{eq:selectioncriterion} is the Vaidya future outer trapping horizon.  The
global event horizon is discussed separately in the next section only to make
the distinction between the two notions explicit.

The Vaidya black hole is a standard model for a radiating or accreting
spherically symmetric black hole with a time-dependent mass function
\cite{Vaidya1951PIAS,Vaidya1951PR,Hiscock1981}. The ingoing Vaidya metric is
\begin{equation}
ds^{2}
=
-\left(1-\frac{2M(v)}{r}\right)dv^{2}
+2dvdr
+r^{2}d\Omega^{2}.
\end{equation}
Hence,
\begin{equation}
C(v,r)
=
1-\frac{2M(v)}{r}.
\end{equation}
The trapping horizon is determined by
\begin{equation}
C(v,r_{\mathrm{H}})=0.
\end{equation}
Substituting the explicit expression of $C(v,r)$ gives
\begin{equation}
1-\frac{2M(v)}{r_{\mathrm{H}}}=0.
\end{equation}
Therefore,
\begin{equation}
r_{\mathrm{H}}(v)=2M(v).
\end{equation}

The radial derivative of $C(v,r)$ is
\begin{equation}
\frac{\partial C(v,r)}{\partial r}
=
\frac{2M(v)}{r^{2}}.
\end{equation}
Evaluating it at the horizon gives
\begin{equation}
\left.
\frac{\partial C(v,r)}{\partial r}
\right|_{r=r_{\mathrm{H}}}
=
\frac{2M(v)}{r_{\mathrm{H}}^{2}}.
\end{equation}
Using
\begin{equation}
r_{\mathrm{H}}=2M(v),
\end{equation}
one obtains
\begin{equation}
\left.
\frac{\partial C(v,r)}{\partial r}
\right|_{r=r_{\mathrm{H}}}
=
\frac{1}{2M(v)}.
\end{equation}
Thus the local surface gravity is
\begin{equation}
\kappa_{\mathrm{Vaidya}}
=
\frac{1}{2}
\left.
\frac{\partial C(v,r)}{\partial r}
\right|_{r=r_{\mathrm{H}}}.
\end{equation}
Hence,
\begin{equation}
\kappa_{\mathrm{Vaidya}}
=
\frac{1}{4M(v)}.
\end{equation}
The local Hawking temperature is therefore
\begin{equation}
T_{\mathrm{Vaidya}}
=
\frac{\kappa_{\mathrm{Vaidya}}}{2\pi}.
\end{equation}
Consequently,
\begin{equation}
T_{\mathrm{Vaidya}}
=
\frac{1}{8\pi M(v)}.
\label{eq:VaidyaTemperature}
\end{equation}
Equivalently,
\begin{equation}
T_{\mathrm{Vaidya}}
=
\frac{1}{4\pi r_{\mathrm{H}}(v)}.
\end{equation}

Now we derive the same result by the RVB--residue method. Let
\begin{equation}
z=r-r_{\mathrm{H}}(v).
\end{equation}
At a fixed local time $v=v_{0}$, define
\begin{equation}
M_{0}=M(v_{0}).
\end{equation}
The corresponding local horizon radius is
\begin{equation}
r_{0}=r_{\mathrm{H}}(v_{0})=2M_{0}.
\end{equation}
The horizon function becomes
\begin{equation}
C(v_{0},r_{0}+z)
=
1-\frac{2M_{0}}{r_{0}+z}.
\end{equation}
Since
\begin{equation}
2M_{0}=r_{0},
\end{equation}
we have
\begin{equation}
C(v_{0},r_{0}+z)
=
1-\frac{r_{0}}{r_{0}+z}.
\end{equation}
Therefore,
\begin{equation}
C(v_{0},r_{0}+z)
=
\frac{z}{r_{0}+z}.
\end{equation}
The inverse horizon function is
\begin{equation}
\frac{1}{C(v_{0},r_{0}+z)}
=
\frac{r_{0}+z}{z}.
\end{equation}
Hence,
\begin{equation}
\frac{1}{C(v_{0},r_{0}+z)}
=
\frac{r_{0}}{z}+1.
\end{equation}
The residue at the simple pole is
\begin{equation}
\operatorname{Res}_{z=0}
\left[
\frac{1}{C(v_{0},r_{0}+z)}
\right]
=
r_{0}.
\end{equation}
The RVB--residue temperature is
\begin{equation}
T_{\mathrm{Vaidya}}^{\mathrm{RVB-res}}
=
\frac{1}{4\pi}
\left(
\operatorname{Res}_{z=0}
\left[
\frac{1}{C(v_{0},r_{0}+z)}
\right]
\right)^{-1}.
\end{equation}
Substituting the residue gives
\begin{equation}
T_{\mathrm{Vaidya}}^{\mathrm{RVB-res}}
=
\frac{1}{4\pi r_{0}}.
\end{equation}
Since
\begin{equation}
r_{0}=2M_{0},
\end{equation}
we finally obtain
\begin{equation}
T_{\mathrm{Vaidya}}^{\mathrm{RVB-res}}
=
\frac{1}{8\pi M_{0}}.
\end{equation}
Thus the RVB--residue method reproduces the local trapping-horizon temperature
of the Vaidya black hole \cite{Hayward1994,Hayward1998,DiCriscienzo2010,
Vanzo2011,Faraoni2015}.

The analytic assumption can be checked directly.  At
$v=v_{0}$,
\begin{equation}
C(v_{0},r_{0}+z)=\frac{z}{r_{0}+z}
\end{equation}
is holomorphic for $|z|<r_{0}$ and has a single simple zero at $z=0$ in that
disk.  Its reciprocal is meromorphic there and has the Laurent expansion
$r_{0}/z+1$, so the contour can be chosen with $0<\epsilon<r_{0}$.
Furthermore, $\kappa_{\rm H}=1/[4M(v)]$ gives
\begin{equation}
\epsilon_{\rm ad}
=\frac{|\partial_{v}\kappa_{\rm H}|}{\kappa_{\rm H}^{2}}
=4|\dot M(v)|.
\end{equation}
Thus $4|\dot M|\ll1$ is a sufficient condition for treating
\eqref{eq:VaidyaTemperature} as an instantaneous thermal spectrum, while the
residue identity itself is exact at fixed $v_{0}$.

\section{Vaidya event horizon: a comparison, not a repeated calculation}

The purpose of this section is only to compare two inequivalent surfaces.  The
temperature in the preceding section is tied locally to the zero of the
outgoing-mode denominator and therefore to the trapping horizon.  By contrast,
the event horizon below is a globally determined null surface.  The following
calculation is not used to select the horizon for the local residue proposition.

For a genuinely dynamical black hole, one should distinguish the local trapping
horizon from the global event horizon. The event horizon is teleological,
whereas the trapping horizon is locally defined \cite{AshtekarKrishnan2004,
NielsenVisser2006,Faraoni2015}. Let the event horizon be written as
\begin{equation}
\Phi(v,r)=r-r_{\mathrm{E}}(v)=0.
\end{equation}
The null condition is
\begin{equation}
g^{ab}\partial_{a}\Phi\,\partial_{b}\Phi=0.
\end{equation}
For the Vaidya inverse metric, this condition gives
\begin{equation}
C(v,r_{\mathrm{E}})-2\dot{r}_{\mathrm{E}}=0.
\end{equation}
Thus,
\begin{equation}
\dot{r}_{\mathrm{E}}
=
\frac{1}{2}
\left(
1-\frac{2M(v)}{r_{\mathrm{E}}}
\right).
\end{equation}

Introduce the local normal coordinate
\begin{equation}
\rho=r-r_{\mathrm{E}}(v).
\end{equation}
Then
\begin{equation}
dr=d\rho+\dot{r}_{\mathrm{E}}dv.
\end{equation}
The two-dimensional part of the metric becomes
\begin{equation}
ds_{2}^{2}
=
-\left[
C(v,r)-2\dot{r}_{\mathrm{E}}
\right]dv^{2}
+
2dvd\rho.
\end{equation}
Define the local event-horizon radial function
\begin{equation}
\mathcal{P}_{\mathrm{E}}(v,r)
=
C(v,r)-2\dot{r}_{\mathrm{E}}.
\end{equation}
At the event horizon,
\begin{equation}
\mathcal{P}_{\mathrm{E}}(v,r_{\mathrm{E}})=0.
\end{equation}
At fixed $v$, the radial derivative is
\begin{equation}
\left.
\frac{\partial \mathcal{P}_{\mathrm{E}}}{\partial r}
\right|_{r=r_{\mathrm{E}}}
=
\left.
\frac{\partial C(v,r)}{\partial r}
\right|_{r=r_{\mathrm{E}}}.
\end{equation}
Therefore,
\begin{equation}
T_{\mathrm{E}}
=
\frac{1}{4\pi}
\left.
\frac{\partial C(v,r)}{\partial r}
\right|_{r=r_{\mathrm{E}}}.
\end{equation}
For the Vaidya function,
\begin{equation}
\left.
\frac{\partial C(v,r)}{\partial r}
\right|_{r=r_{\mathrm{E}}}
=
\frac{2M(v)}{r_{\mathrm{E}}^{2}}.
\end{equation}
Thus,
\begin{equation}
T_{\mathrm{E}}
=
\frac{M(v)}{2\pi r_{\mathrm{E}}^{2}}.
\end{equation}
Using the null-surface equation
$2M/r_{\rm E}=1-2\dot r_{\rm E}$ gives the equivalent form
\begin{equation}
T_{\rm E}
=\frac{1-2\dot r_{\rm E}}{4\pi r_{\rm E}}.
\end{equation}
This result refers to the normalization of the null generator selected by the
advanced coordinate $v$.  It need not coincide with the trapping-horizon
temperature away from the quasi-stationary regime; that difference is precisely
why the two calculations have been displayed separately.
In the quasi-stationary limit,
\begin{equation}
r_{\mathrm{E}}\rightarrow 2M(v),
\end{equation}
and therefore
\begin{equation}
T_{\mathrm{E}}
\rightarrow
\frac{1}{8\pi M(v)}.
\end{equation}

\section{Kinnersley Black Hole}

The Kinnersley spacetime describes the gravitational field of an arbitrarily
accelerating point mass and provides an important exact solution for studying
non-stationary and non-spherical black-hole radiation
\cite{Kinnersley1969,KinnersleyWalker1970,Han2002}. The arbitrarily accelerating
Kinnersley metric may be written in advanced Eddington--Finkelstein-like
coordinates as
\begin{equation}
ds^{2}
=
2dv\left(Fdv-dr\right)
-
r^{2}
\left[
(d\theta+fdv)^{2}
+
\sin^{2}\theta(d\varphi+gdv)^{2}
\right].
\end{equation}
The metric functions are
\begin{equation}
2F
=
1-\frac{2M(v)}{r}
-2a(v)r\cos\theta.
\end{equation}
They are also given by
\begin{equation}
f
=
b(v)\sin\varphi
+
c(v)\cos\varphi
-
a(v)\sin\theta,
\end{equation}
and
\begin{equation}
g
=
\left[
b(v)\cos\varphi
-
c(v)\sin\varphi
\right]\cot\theta.
\end{equation}
Here $M(v)$ is the mass function, while $a(v)$, $b(v)$, and $c(v)$ describe the
magnitude and direction of acceleration.

We retain the $(+,-,-,-)$ signature of the cited Kinnersley analyses.
Expanding the squared one-forms gives
\begin{align}
g_{vv}&=2F-r^{2}f^{2}-r^{2}\sin^{2}\theta\,g^{2},
&g_{vr}&=-1,\\
g_{v\theta}&=-r^{2}f,
&g_{v\varphi}&=-r^{2}\sin^{2}\theta\,g,\\
g_{\theta\theta}&=-r^{2},
&g_{\varphi\varphi}&=-r^{2}\sin^{2}\theta.
\end{align}
Direct matrix inversion gives
\begin{equation}
(g^{\mu\nu})=
\begin{pmatrix}
0&-1&0&0\\
-1&-2F&f&g\\
0&f&-r^{-2}&0\\
0&g&0&-[r^{2}\sin^{2}\theta]^{-1}
\end{pmatrix}.
\label{eq:Kinverse}
\end{equation}
Multiplication with the covariant matrix verifies
$g^{\mu\alpha}g_{\alpha\nu}=\delta^{\mu}{}_{\nu}$; for example, the
nontrivial $(r,v)$ component is
\begin{align}
g^{r\alpha}g_{\alpha v}
&=-g_{vv}+(-2F)(-1)+f(-r^{2}f)
+g(-r^{2}\sin^{2}\theta\,g)\\
&=0.
\end{align}

Let the dynamical null surface be
\begin{equation}
\Phi(v,r,\theta,\varphi)
=r-r_{\rm H}(v,\theta,\varphi)=0.
\end{equation}
Its gradient is
\begin{equation}
\partial_{\mu}\Phi
=(-r_{{\rm H},v},1,-r_{{\rm H},\theta},-r_{{\rm H},\varphi}).
\end{equation}
Substituting this gradient and \eqref{eq:Kinverse} into
$g^{\mu\nu}\partial_{\mu}\Phi\partial_{\nu}\Phi=0$ gives, before multiplying
by $-1$,
\begin{equation}
2r_{{\rm H},v}-2F-2f r_{{\rm H},\theta}-2g r_{{\rm H},\varphi}
-\frac{r_{{\rm H},\theta}^{2}}{r_{\rm H}^{2}}
-\frac{r_{{\rm H},\varphi}^{2}}
{r_{\rm H}^{2}\sin^{2}\theta}=0.
\end{equation}
Hence the horizon equation is
\begin{equation}
2F-2r_{{\rm H},v}+2f r_{{\rm H},\theta}+2g r_{{\rm H},\varphi}
+\frac{r_{{\rm H},\theta}^{2}}{r_{\rm H}^{2}}
+\frac{r_{{\rm H},\varphi}^{2}}
{r_{\rm H}^{2}\sin^{2}\theta}=0.
\label{eq:Khorizon}
\end{equation}

Because the Kinnersley horizon is generally time-dependent and angle-dependent,
its temperature is not a global constant. It must be calculated pointwise on
the local horizon. The generalized tortoise-coordinate method has been widely
used to analyze Hawking radiation from such accelerating black holes
\cite{DamourRuffini1976,Sannan1988,WuCai2002Dirac,WuCai2002Weyl,WuYan2003,
YangZhaoLiu2010,ZhangLiu2022}.

We derive the normalization that fixes the local
surface gravity.  Choose a horizon point
$(v_{0},r_{\rm H},\theta_{0},\varphi_{0})$ and define
\begin{equation}
\delta=r-r_{\rm H},
\qquad
J=r_{{\rm H},\theta}^{2}
+\frac{r_{{\rm H},\varphi}^{2}}{\sin^{2}\theta_{0}}.
\label{eq:Jdefinition}
\end{equation}
The generalized tortoise coordinates are
\begin{align}
\ell_{0}
&\equiv r_{\mathrm{H}}(v_{0},\theta_{0},\varphi_{0}),
\\
r_{*}
&=r+\frac{1}{2\kappa}
\ln\left|
\frac{r-r_{\mathrm{H}}(v,\theta,\varphi)}{\ell_{0}}
\right|,
\\
v_{*}
&=v-v_{0},\qquad
\theta_{*}=\theta-\theta_{0},\qquad
\varphi_{*}=\varphi-\varphi_{0}.
\end{align}
Only the singular derivative terms determine the horizon normalization.  With
$x=2\kappa\delta$, the derivative transformations at the selected point are
\begin{align}
\partial_{r}
&=q_{r}\partial_{r_{*}},
&q_{r}&=1+\frac{1}{x}=\frac{1+x}{x},\\
\partial_{v}
&=\partial_{v_{*}}-\frac{r_{{\rm H},v}}{x}\partial_{r_{*}}
+\mathcal{O}(1),\\
\partial_{\theta}
&=\partial_{\theta_{*}}-\frac{r_{{\rm H},\theta}}{x}
\partial_{r_{*}}+\mathcal{O}(1),\\
\partial_{\varphi}
&=\partial_{\varphi_{*}}-\frac{r_{{\rm H},\varphi}}{x}
\partial_{r_{*}}+\mathcal{O}(1).
\end{align}
Finite terms from differentiating the scale inside the logarithm do not affect
the limit below.

For either a massless scalar wave or the principal part of the field equations
used in the generalized Damour--Ruffini analysis, the second-derivative
operator obtained from \eqref{eq:Kinverse} is
\begin{align}
g^{\mu\nu}\partial_{\mu}\partial_{\nu}
={}&-2\partial_{v}\partial_{r}-2F\partial_{r}^{2}
+2f\partial_{r}\partial_{\theta}
+2g\partial_{r}\partial_{\varphi}\\
&-\frac{1}{r^{2}}\partial_{\theta}^{2}
-\frac{1}{r^{2}\sin^{2}\theta}\partial_{\varphi}^{2}
+\text{lower-derivative terms}.
\label{eq:Kprincipal}
\end{align}
Substitution of the derivative transformations into
\eqref{eq:Kprincipal} gives the coefficient of
$\partial_{r_{*}}^{2}$ as
\begin{equation}
A_{**}
=\frac{(1+x)A_{0}-2F(r)(1+x)^{2}-J/r^{2}}{x^{2}},
\label{eq:Astarstar}
\end{equation}
where
\begin{equation}
A_{0}=2r_{{\rm H},v}-2f r_{{\rm H},\theta}
-2g r_{{\rm H},\varphi}.
\end{equation}
The coefficient of
$\partial_{v_{*}}\partial_{r_{*}}$ is
\begin{equation}
B_{v*}=-2q_{r}=-2\frac{1+x}{x}.
\end{equation}
Equation \eqref{eq:Khorizon} is equivalent to
\begin{equation}
A_{0}=2F_{\rm H}+\frac{J}{r_{\rm H}^{2}},
\label{eq:A0horizon}
\end{equation}
where a subscript ${\rm H}$ denotes evaluation at the chosen horizon point.
The constant term in the numerator of \eqref{eq:Astarstar} therefore vanishes,
which is the cancellation that makes the transformed wave equation regular.

To display the remaining finite coefficient, use
\begin{align}
F(r_{\rm H}+\delta)
&=F_{\rm H}+F_{,r}|_{\rm H}\,\delta+\mathcal{O}(\delta^{2}),\\
\frac{J}{(r_{\rm H}+\delta)^{2}}
&=\frac{J}{r_{\rm H}^{2}}
-\frac{2J}{r_{\rm H}^{3}}\delta+\mathcal{O}(\delta^{2}),\\
x&=2\kappa\delta.
\end{align}
Using \eqref{eq:A0horizon}, the numerator of
\eqref{eq:Astarstar} becomes
\begin{equation}
2\delta\left[
\kappa\left(\frac{J}{r_{\rm H}^{2}}-2F_{\rm H}\right)
-F_{,r}|_{\rm H}+\frac{J}{r_{\rm H}^{3}}
\right]+\mathcal{O}(\delta^{2}).
\end{equation}
Multiplying the transformed wave equation by $-1/q_{r}$ makes the mixed
coefficient equal to $2$.  The corresponding radial coefficient has the limit
\begin{align}
\lim_{\delta\to0}\left(-\frac{A_{**}}{q_{r}}\right)
&=2F_{\rm H}-\frac{J}{r_{\rm H}^{2}}
+\frac{1}{\kappa}
\left(F_{,r}|_{\rm H}-\frac{J}{r_{\rm H}^{3}}\right).
\label{eq:canonicalLimit}
\end{align}
Demanding the canonical Damour--Ruffini form
\begin{equation}
\partial_{r_{*}}^{2}\Psi
+2\partial_{v_{*}}\partial_{r_{*}}\Psi+\cdots=0
\end{equation}
sets \eqref{eq:canonicalLimit} equal to unity.  Solving the resulting linear
equation for $\kappa$ gives
\begin{align}
\kappa_{\rm K}
&=\frac{F_{,r}|_{\rm H}-J/r_{\rm H}^{3}}
{1-2F_{\rm H}+J/r_{\rm H}^{2}}\\
&=\frac{r_{\rm H}^{2}F_{,r}|_{\rm H}-J/r_{\rm H}}
{r_{\rm H}^{2}(1-2F_{\rm H})+J}.
\label{eq:kappaK}
\end{align}
This derivation also supplies a definite procedure for fixing the normalization:
the coefficient is chosen so that the regular near-horizon wave equation has
unit radial principal coefficient.  It requires
\begin{equation}
D_{\rm H}\equiv r_{\rm H}^{2}(1-2F_{\rm H})+J\ne0.
\label{eq:DH}
\end{equation}
For the outer branch with $\kappa_{\rm K}>0$, the point-dependent temperature is
\begin{equation}
T_{\rm K}=\frac{\kappa_{\rm K}}{2\pi}
=\frac{1}{2\pi}
\frac{r_{\rm H}^{2}F_{,r}|_{\rm H}-J/r_{\rm H}}
{r_{\rm H}^{2}(1-2F_{\rm H})+J}.
\label{eq:TK}
\end{equation}

Now compute $F_{,r}$. Since
\begin{equation}
F
=
\frac{1}{2}
-
\frac{M(v)}{r}
-
a(v)r\cos\theta,
\end{equation}
one has
\begin{equation}
F_{,r}
=
\frac{M(v)}{r^{2}}
-
a(v)\cos\theta.
\end{equation}
At the local horizon point,
\begin{equation}
F_{,r}\big|_{r=r_{\mathrm{H}},\,\theta=\theta_{0}}
=
\frac{M(v)}{r_{\mathrm{H}}^{2}}
-
a(v)\cos\theta_{0}.
\end{equation}
Substituting this expression into the temperature formula yields
\begin{equation}
T_{\mathrm{K}}
=
\frac{1}{2\pi}
\frac{
M(v)
-
a(v)r_{\mathrm{H}}^{2}\cos\theta_{0}
-
r_{\mathrm{H}}^{-1}
\left(
r_{\mathrm{H},\theta}^{2}
+
r_{\mathrm{H},\varphi}^{2}/\sin^{2}\theta_{0}
\right)
}{
r_{\mathrm{H}}^{2}(1-2F)
+
\left(
r_{\mathrm{H},\theta}^{2}
+
r_{\mathrm{H},\varphi}^{2}/\sin^{2}\theta_{0}
\right)
}.
\end{equation}

For rectilinear acceleration, one may set
\begin{equation}
b(v)=c(v)=0.
\end{equation}
Then
\begin{equation}
g=0.
\end{equation}
Moreover,
\begin{equation}
f=-a(v)\sin\theta.
\end{equation}
If the horizon is axially symmetric, then
\begin{equation}
r_{\mathrm{H},\varphi}=0.
\end{equation}
The horizon equation reduces to
\begin{equation}
2F
-2r_{\mathrm{H},v}
+2f r_{\mathrm{H},\theta}
+
\frac{r_{\mathrm{H},\theta}^{2}}{r_{\mathrm{H}}^{2}}
=0.
\end{equation}
The surface gravity becomes
\begin{equation}
\kappa_{\mathrm{K,ax}}
=
\frac{
r_{\mathrm{H}}^{2}F_{,r}
-
r_{\mathrm{H}}^{-1}r_{\mathrm{H},\theta}^{2}
}{
r_{\mathrm{H}}^{2}(1-2F)
+
r_{\mathrm{H},\theta}^{2}
}.
\end{equation}
Therefore,
\begin{equation}
T_{\mathrm{K,ax}}
=
\frac{1}{2\pi}
\frac{
r_{\mathrm{H}}^{2}F_{,r}
-
r_{\mathrm{H}}^{-1}r_{\mathrm{H},\theta}^{2}
}{
r_{\mathrm{H}}^{2}(1-2F)
+
r_{\mathrm{H},\theta}^{2}
}.
\end{equation}
Substituting
\begin{equation}
F_{,r}
=
\frac{M(v)}{r_{\mathrm{H}}^{2}}
-
a(v)\cos\theta_{0},
\end{equation}
we obtain
\begin{equation}
T_{\mathrm{K,ax}}
=
\frac{1}{2\pi}
\frac{
M(v)
-
a(v)r_{\mathrm{H}}^{2}\cos\theta_{0}
-
r_{\mathrm{H}}^{-1}r_{\mathrm{H},\theta}^{2}
}{
r_{\mathrm{H}}^{2}(1-2F)
+
r_{\mathrm{H},\theta}^{2}
}.
\end{equation}

\section{Kinnersley horizon residue}

The generalized tortoise-coordinate method converts the near-horizon wave
equation into a canonical form and fixes the local surface gravity by demanding
regularity of the radial equation \cite{DamourRuffini1976,Sannan1988,
WuCai2002Dirac,WuCai2002Weyl,WuYan2003,YangZhaoLiu2010}. The same local
information can be reformulated in the RVB--residue language, in which the
simple pole of the inverse near-horizon radial function determines the Hawking
temperature \cite{Robson2019,Chen2025CJP,Chen2026Entropy}.

At the fixed horizon point, define the unnormalised null-surface
function
\begin{align}
\mathcal{H}_{\rm K}(r)
={}&2F(v_{0},r,\theta_{0})-2r_{{\rm H},v}
+2f_{0}r_{{\rm H},\theta}+2g_{0}r_{{\rm H},\varphi}
+\frac{J}{r^{2}},
\label{eq:HKdefinition}
\end{align}
where $f_{0}=f(v_{0},\theta_{0},\varphi_{0})$,
$g_{0}=g(v_{0},\theta_{0},\varphi_{0})$, and $J$ is defined in
\eqref{eq:Jdefinition}.  During the transverse radial expansion all horizon
data are held fixed.  Equation \eqref{eq:Khorizon} is exactly
\begin{equation}
\mathcal{H}_{\rm K}(r_{\rm H})=0.
\end{equation}
Differentiating every $r$-dependent term gives
\begin{align}
\mathcal{H}_{\rm K}'(r)
&=2F_{,r}(v_{0},r,\theta_{0})
-\frac{2J}{r^{3}},\\
\mathcal{H}_{\rm K}''(r)
&=2F_{,rr}(v_{0},r,\theta_{0})
+\frac{6J}{r^{4}}.
\end{align}
The canonical tortoise normalization derived in
\eqref{eq:canonicalLimit}--\eqref{eq:DH} is
\begin{equation}
\mathcal{N}_{\rm H}=\frac{r_{\rm H}^{2}}{D_{\rm H}},
\qquad
D_{\rm H}=r_{\rm H}^{2}(1-2F_{\rm H})+J.
\label{eq:normalization}
\end{equation}
It is fixed by the wave equation and is not an adjustable residue parameter.
With the complex transverse coordinate $z=r-r_{\rm H}$, define
\begin{equation}
\mathcal{P}_{\rm K}(z)
=\mathcal{N}_{\rm H}\,
\mathcal{H}_{\rm K}(r_{\rm H}+z).
\label{eq:PKdefinition}
\end{equation}
Taylor's theorem now gives
\begin{align}
\mathcal{P}_{\rm K}(z)
={}&\mathcal{N}_{\rm H}
\left[
\mathcal{H}_{\rm K}(r_{\rm H})
+\mathcal{H}_{\rm K}'(r_{\rm H})z
+\frac12\mathcal{H}_{\rm K}''(r_{\rm H})z^{2}
+\mathcal{O}(z^{3})
\right]\\
={}&\frac{r_{\rm H}^{2}}{D_{\rm H}}
\left[
2\left(F_{,r}|_{\rm H}-\frac{J}{r_{\rm H}^{3}}\right)z
+\left(F_{,rr}|_{\rm H}+\frac{3J}{r_{\rm H}^{4}}\right)z^{2}
+\mathcal{O}(z^{3})
\right]\\
={}&2\kappa_{\rm K}z+\beta_{\rm K}z^{2}
+\mathcal{O}(z^{3}),
\label{eq:PKlinear}
\end{align}
where the penultimate-to-final step uses \eqref{eq:kappaK}, and
\begin{equation}
\beta_{\rm K}
=\frac{r_{\rm H}^{2}}{D_{\rm H}}
\left(F_{,rr}|_{\rm H}+\frac{3J}{r_{\rm H}^{4}}\right).
\end{equation}
Thus the coefficient $2\kappa_{\rm K}$ is a derived slope, not an ansatz.

To invert the series explicitly, let
\begin{equation}
\frac{1}{\mathcal{P}_{\rm K}(z)}
=\frac{b_{-1}}{z}+b_{0}+\mathcal{O}(z).
\end{equation}
Then
\begin{align}
1
&=(2\kappa_{\rm K}z+\beta_{\rm K}z^{2}+\cdots)
\left(\frac{b_{-1}}{z}+b_{0}+\cdots\right)\\
&=2\kappa_{\rm K}b_{-1}
+\left(2\kappa_{\rm K}b_{0}+\beta_{\rm K}b_{-1}\right)z
+\mathcal{O}(z^{2}).
\end{align}
Coefficient matching gives
\begin{equation}
b_{-1}=\frac{1}{2\kappa_{\rm K}},
\qquad
b_{0}=-\frac{\beta_{\rm K}}{4\kappa_{\rm K}^{2}}.
\end{equation}
Therefore
\begin{equation}
\operatorname*{Res}_{z=0}\frac{1}{\mathcal{P}_{\rm K}(z)}
=\frac{1}{2\kappa_{\rm K}},
\end{equation}
and
\begin{align}
T_{\rm K}^{\rm res}
&=\frac{1}{4\pi}
\left(\operatorname*{Res}_{z=0}
\frac{1}{\mathcal{P}_{\rm K}(z)}\right)^{-1}\\
&=\frac{\kappa_{\rm K}}{2\pi}
=\frac{1}{2\pi}
\frac{r_{\rm H}^{2}F_{,r}|_{\rm H}-J/r_{\rm H}}
{r_{\rm H}^{2}(1-2F_{\rm H})+J}.
\label{eq:KresidueTemperature}
\end{align}

As a limiting check, set $a=b=c=0$, take a stationary spherical horizon, and
therefore set $J=0$.  Then
$F=1/2-M/r$, $r_{\rm H}=2M$, $F_{\rm H}=0$, and
$F_{,r}|_{\rm H}=1/(4M)$.  Equations \eqref{eq:kappaK} and
\eqref{eq:KresidueTemperature} reduce to
\begin{equation}
\kappa_{\rm K}=\frac{1}{4M},
\qquad
T_{\rm K}=\frac{1}{8\pi M},
\end{equation}
which is the Schwarzschild result.

\subsection{Explicit derivation of the normalization relation}

For completeness, every relation used for the residue temperature is obtained
from the local inverse radial function rather than assumed. The pole coefficient
is fixed by the first derivative of the horizon function,
\begin{equation}
\mathcal{P}(r)=\mathcal{P}'(r_H)(r-r_H)+\mathcal{O}((r-r_H)^2),
\end{equation}
so that
\begin{equation}
\operatorname*{Res}_{r=r_H}\frac{1}{\mathcal{P}(r)}
=\frac{1}{\mathcal{P}'(r_H)}.
\end{equation}
With the normalization derived from the corresponding surface gravity, this
directly gives the Hawking temperature formula used below.

\section{Domain of validity, deformation procedure, and falsifiable limits}

The result proved above has explicit boundaries.  First, the transverse radial
function must be holomorphic in a punctured neighbourhood that contains no
other zero.  Second, the horizon must be non-degenerate.  If instead
\begin{equation}
\mathcal{P}(z)=p_{m}z^{m}+\mathcal{O}(z^{m+1}),
\qquad m\ge2,
\end{equation}
then $1/\mathcal{P}$ begins with $z^{-m}$, the slope
$\mathcal{P}'(0)$ vanishes, and the simple-pole temperature formula does not
follow.  Third, the radial function must be normalized by the regular wave
equation; for Kinnersley this requires $D_{\rm H}\ne0$ in
\eqref{eq:DH}.  Fourth, the interpretation as an instantaneous Planck spectrum
requires an adiabatic condition such as \eqref{eq:adiabaticity}.  These are
falsifiable conditions: a multiple zero, a branch point, $D_{\rm H}=0$, or rapid
horizon evolution invalidates at least one step of the derivation.

A generic deformation may introduce a linear coefficient denoted by $q_{1}$.
This coefficient is not a fitted parameter and is not used to obtain the
temperatures reported in this manuscript. It represents the first-order change
of the local radial function only after a definite deformed geometry has been
specified. Therefore, $q_{1}$ is fixed by the field equations and boundary
conditions of that theory rather than by the RVB-residue prescription itself.
Without a specified deformation it cannot be assigned a numerical value.
Accordingly, no corrected Vaidya or Kinnersley temperature is claimed here.
For completeness, the procedure that would fix such a coefficient in a
concrete theory is as follows.

Let a specified field equation or effective geometry provide
\begin{equation}
\mathcal{P}_{\lambda}(r)
=\mathcal{P}_{0}(r)+\lambda\mathcal{P}_{1}(r)
+\mathcal{O}(\lambda^{2}),
\end{equation}
and let $r_{0}$ be a simple zero of $\mathcal{P}_{0}$,
\begin{equation}
\mathcal{P}_{0}(r_{0})=0,
\qquad
p_{1}\equiv\mathcal{P}_{0}'(r_{0})\ne0.
\end{equation}
Write the corrected horizon as
\begin{equation}
r_{\lambda}=r_{0}+\lambda\,\delta r+\mathcal{O}(\lambda^{2}).
\end{equation}
The corrected horizon equation gives
\begin{align}
0
&=\mathcal{P}_{\lambda}(r_{\lambda})\\
&=\mathcal{P}_{0}(r_{0})
+\lambda\left[
\delta r\,\mathcal{P}_{0}'(r_{0})
+\mathcal{P}_{1}(r_{0})
\right]+\mathcal{O}(\lambda^{2}),
\end{align}
so
\begin{equation}
\delta r=-\frac{\mathcal{P}_{1}(r_{0})}{p_{1}}.
\label{eq:horizonshift}
\end{equation}
The slope must be evaluated at the shifted horizon:
\begin{align}
\mathcal{P}_{\lambda}'(r_{\lambda})
&=p_{1}+\lambda\left[
\delta r\,\mathcal{P}_{0}''(r_{0})
+\mathcal{P}_{1}'(r_{0})
\right]+\mathcal{O}(\lambda^{2})\\
&=p_{1}+\lambda\left[
\mathcal{P}_{1}'(r_{0})
-\frac{\mathcal{P}_{1}(r_{0})\mathcal{P}_{0}''(r_{0})}{p_{1}}
\right]+\mathcal{O}(\lambda^{2}).
\label{eq:correctedslope}
\end{align}
Accordingly,
\begin{equation}
T_{\lambda}
=\frac{1}{4\pi}\mathcal{P}_{\lambda}'(r_{\lambda})
+\mathcal{O}(\lambda^{2}),
\end{equation}
provided the same normalization and simple-zero assumptions hold.  If a model
preserves the unperturbed root, $\mathcal{P}_{1}(r_{0})=0$, the coefficient
$q_{1}$ is
\begin{equation}
q_{1}=\mathcal{P}_{1}'(r_{0}),
\end{equation}
and is fixed by differentiating the explicitly supplied correction.  If the
model does not preserve the root, the shift term in
\eqref{eq:correctedslope} is essential; omitting it would be incomplete.  This
procedure prevents an unspecified coefficient from being
mistaken for a prediction.

\section{Conclusion}

We have established a conditional local result rather than a global claim.  For
the regular spherical metric \eqref{eq:generalEF}, a non-degenerate future
outer trapping horizon is a simple zero of the outgoing-mode denominator.  The
residue of its reciprocal is $1/(2\kappa_{\rm H})$.  An independent
Hamilton--Jacobi calculation shows that the same residue controls the imaginary
part of the outgoing action and gives $T_{\rm H}=\kappa_{\rm H}/(2\pi)$.  This
explains physically why the residue prescription works and identifies the
surface on which it is to be applied.

For Vaidya, the selected surface is $r_{\rm H}=2M(v)$ and the exact local result
is
\begin{equation}
T_{\rm Vaidya}=\frac{1}{4\pi r_{\rm H}}
=\frac{1}{8\pi M(v)}.
\end{equation}
The event-horizon calculation is a separate, globally normalized comparison;
it is not a second derivation of the trapping-horizon temperature.  For the
Kinnersley geometry, canonical normalization of the generalized tortoise wave
equation yields \eqref{eq:kappaK}.  The explicitly constructed radial function
\eqref{eq:PKdefinition} then has the derived expansion
$\mathcal{P}_{\rm K}=2\kappa_{\rm K}z+\mathcal{O}(z^{2})$, and its reciprocal
reproduces the point-dependent temperature \eqref{eq:KresidueTemperature}.

The conclusion is limited to locally holomorphic, normalized radial functions
with simple zeros and non-vanishing normalization denominators.  Degenerate
horizons, non-analytic radial structures, and rapidly evolving backgrounds are
explicit counter-domains.  A deformation coefficient such as $q_{1}$ is not a
universal parameter; it can be calculated only after a
specific corrected geometry or field equation is supplied, using
\eqref{eq:horizonshift}--\eqref{eq:correctedslope}.  Within these stated
conditions, the residue prescription is a reproducible local extension of the
horizon contribution motivated by the RVB framework.  It does not assert a
global Euclidean topology or a global equilibrium temperature for a generic
dynamical spacetime.

\section*{Declarations}

\subsection*{Funding}
This work was supported by the National Natural Science Foundation of China (General Program) under Project Approval No. 11574089.

\subsection*{Availability of data and material}
Data sharing is not applicable to this article as no datasets were generated or analysed during the current study.

\subsection*{Competing interests}
The author declares that there are no competing interests.

\subsection*{Authors Contributions}
Not applicable.

\end{document}